\pdfoutput=1

\documentclass[11pt,twoside,a4paper,cmspaper,final,collab]{cms-tdr}
\def\svnVersion{171944}\def\cmsCernNo{2012-283}\def\cmsCernDate{\today}\def\cmsMessage{Submitted to the Journal of High Energy Physics}

\begin{document}\cmsNoteHeader{SMP-12-009}

\hyphenation{had-ron-i-za-tion}
\hyphenation{cal-or-i-me-ter}
\hyphenation{de-vices}

\RCS$Revision: 152685 $
\RCS$HeadURL: svn+ssh://svn.cern.ch/reps/tdr2/papers/SMP-12-009/trunk/SMP-12-009.tex $
\RCS$Id: SMP-12-009.tex 152685 2012-10-14 20:59:28Z alverson $
\newlength\cmsFigWidth
\ifthenelse{\boolean{cms@external}}{\setlength\cmsFigWidth{0.85\columnwidth}}{\setlength\cmsFigWidth{0.4\textwidth}}
\ifthenelse{\boolean{cms@external}}{\providecommand{\cmsLeft}{top}}{\providecommand{\cmsLeft}{left}}
\ifthenelse{\boolean{cms@external}}{\providecommand{\cmsRight}{bottom}}{\providecommand{\cmsRight}{right}}

\newcommand{\Zee}{\ensuremath{\cPZ\to\Pep\Pem}}
\newcommand{\nanob}{\mbox{{\rm ~nb}~}}
\newcommand{\pp}{\ensuremath{\Pp\Pp}}
\newcommand{\fb}{\ensuremath{\mathrm{fb}}}
\newcommand{\pb}{\ensuremath{\mathrm{pb}}}
\newcommand{\ifb}{\ensuremath{\mathrm{fb^{-1}}}}
\newcommand{\ipb}{\ensuremath{\mathrm{pb^{-1}}}}
\newcommand{\Hmu}{\ensuremath{\PH\to\cPZ\cPZ^{(\ast)}\to 4\Pgm}}
\newcommand{\mmu}{\ensuremath{m_{4\Pgm}}}
\newcommand{\mH}{\ensuremath{m_{\PH}}}
\newcommand{\lnQ}{\ensuremath{\mathrm{ln}(Q)}}
\newcommand{\mlnQ}{\ensuremath{-2\,\ln(Q)}}
\newcommand{\LS}{\ensuremath{\mathcal{L}(S)}}
\newcommand{\LB}{\ensuremath{\mathcal{L}(B)}}
\newcommand{\LSB}{\ensuremath{\mathcal{L}(S+B)}}
\newcommand{\mZ}{\ensuremath{m_{\cPZ}}}
\newcommand{\Zbb}{\ensuremath{(\cPZ/\Pgg^*){\cPqb\cPaqb}}}
\newcommand{\ZZ}{\ensuremath{\cPZ/\Pgg^*\cPZ/\Pgg^*}}
\newcommand{\Zgam}{\ensuremath{\cPZ/\Pgg^*}}
\newcommand{\mumu}{\ensuremath{\Pgmp\Pgmm}}

\newcommand{\BR}{$\mathcal{B}(\mathrm{Z} \to 4\ell) = \left( 4.2^{+0.9}_{-0.8}\,\text{(stat. )} \pm 0.2\,\text{(syst.)}   \right) \times 10^{-6}$}
\newcommand{\XSxBR}{$\sigma(\pp\to\mathrm{Z}) \, \mathcal{B}(\mathrm{Z} \to 4\ell) = 112^{+23}_{-20}\,\text{(stat.)}^{+7}_{-5}\,\text{(syst.)}^{+3}_{-2}\,\text{(lumi.)}\fbinv$}
\newcommand{\Nobs}{28}
\newcommand{\nTOTexp}{30.0}
\newcommand{\nTOTexpErr}{$30.0 \pm 2.6$}
\newcommand{\nZexp}{29.4}
\newcommand{\nZexpErr}{$29.4 \pm 2.6$}
\newcommand{\nBKGDexp}{0.56}
\newcommand{\nBKGDexpErr}{$0.6 \pm 0.2$}
\newcommand{\Significance}{9.7$\sigma$}

\cmsNoteHeader{SMP-12-009} 
\title{Observation of $\mathrm{Z}$ decays to four leptons with the CMS detector at the LHC}

\date{\today}

\abstract{
The first observation of the Z boson decaying to four leptons in proton-proton collisions is presented. The analyzed data set corresponds
to an integrated
luminosity of 5.02\fbinv at $\sqrt{s}=7$\TeV
collected by the CMS detector at the Large Hadron Collider.
A pronounced resonance peak, with a statistical significance of 9.7$\,\sigma$,
is observed in the distribution
of the invariant mass of four leptons (electrons and/or muons) with  mass and
width consistent with expectations for Z boson decays.
The branching fraction and cross section reported here are defined by phase space restrictions on the leptons, namely, $80 < m_{4\ell} <100$\GeV, where $m_{4\ell}$ is the invariant mass of the four leptons, and $m_{\ell\ell} > 4$\GeV for all pairs of leptons, where $m_{\ell\ell}$ is the two-lepton invariant mass.
The measured branching fraction is $\mathcal{B}(\mathrm{Z} \to 4\ell) = \left( 4.2^{+0.9}_{-0.8} \text{(stat.)} \pm 0.2 \text{(syst.)}  \right) \times  10^{-6}$ and agrees with the standard model prediction of $4.45 \times 10^{-6}$.  The measured cross section times branching fraction is $\sigma(\pp\to\mathrm{Z}) \,  \mathcal{B}(\mathrm{Z} \to 4\ell) =  112^{+23}_{-20} \text{(stat.)} ^{+7}_{-5} \text{(syst.)} ^{+3}_{-2} \text{(lumi.)}\unit{fb}$, also consistent with the standard model prediction of 120\unit{fb}. The four-lepton mass peak arising from $\mathrm{Z} \to 4\ell$ decays provides a calibration channel for the Higgs boson search in the $\mathrm{H} \to \mathrm{ZZ} \to 4\ell$ decay mode.
}

\hypersetup{%
pdfauthor={CMS Collaboration},%
pdftitle={Observation of Z decays to four leptons with the CMS detector at the LHC},%
pdfsubject={CMS},%
pdfkeywords={CMS, physics, software, computing}}

\maketitle 

\section{Introduction}
We present the first observation of $\mathrm{Z} \to 4\ell$ decays in
\pp ~collisions at $\sqrt{s}=7\TeV$, and measurements of the
production cross section times branching fraction for $\mathrm{Z}$ decays into four leptons $\sigma(\mathrm{pp} \to \mathrm{Z}) \, \mathcal{B}(\mathrm{Z} \to 4\ell)$
and the branching fraction $\mathcal{B}(\mathrm{Z} \to 4\ell)$. In this paper,
$\ell$ denotes an electron or muon. Previously, all four LEP
collaborations reported observations of
final states with four fermions,
$\mathrm{e}^+\mathrm{e}^- \to 4f$, which include $\mathrm{e}^+\mathrm{e}^- \to \mathrm{Z} \to 4f$~\cite{Decamp:1991sh,Acton:1992ip,Abreu:1993mq,Adam:1993xs}.
However, the observation of $\mathrm{Z} \to 4\ell$ decays in \pp ~collisions
is of special interest.
The clean resonant peak in the four-lepton invariant mass distribution
at $m_{4\ell} = m_Z$ can be used for direct calibration of the four-lepton
mass scale, the four-lepton mass resolution, and the overall
four-lepton reconstruction efficiency in phase space similar to
the light Higgs boson four-lepton decays, $\mathrm{H} \to \mathrm{ZZ} \to 4\ell$.
The $\mathrm{pp} \to \mathrm{Z} \to 4\ell$ process
and its implications for the $\mathrm{H} \to \mathrm{ZZ} \to 4\ell$ search at the Large Hadron Collider (LHC) were first
studied in Ref.~\cite{NOTE-2006-057}.

These results are based on the data collected in 2010 and 2011 by the Compact Muon Solenoid (CMS) detector~\cite{Chatrchyan:2008zzk} at the LHC.  The data set used in the analysis
corresponds to a total integrated luminosity of $5.02\pm0.11$\fbinv.

\section{CMS detector}
\label{sec:CMS-detector}

The central feature of the CMS apparatus is a
superconducting solenoid 13\unit{m} in length and 6\unit{m} in diameter providing an axial magnetic field of 3.8\unit{T}. Within the field volume are
the silicon pixel and strip tracker, the crystal electromagnetic
calorimeter (ECAL), and the brass/scintillator hadron calorimeter (HCAL).
Charged particle trajectories are measured by the
tracker, covering $0 < \phi < 2\pi$ in azimuth and $|\eta| < 2.5$,
where the pseudorapidity is defined as
$\eta = -\ln [\tan (\theta/2)]$,
with $\theta$ being the polar angle of the trajectory of the particle
with respect to the anticlockwise beam direction.
Muons are measured in the pseudorapidity range $|\eta|< 2.4$,
with detection planes made using three technologies: Drift Tubes, Cathode
Strip Chambers, and Resistive Plate Chambers.
Matching the muons to the
tracks measured in the silicon tracker results in a transverse momentum
($\pt$) resolution of about 2\% in the muon $\pt$ range relevant for
the analysis presented in this paper (from a few to a few tens of \GeVns).
The ECAL consists of nearly 76\,000 lead tungstate
crystals distributed in the barrel region (EB, $|\eta| < 1.479$) and
two endcap regions (EE, $1.479 < |\eta| < 3$), and
has an ultimate energy resolution better than 0.5\%
for unconverted photons with transverse energies above 100\GeV. The
electron energy resolution is 3\% or better for the range of
energies relevant for the measurement reported in this paper.
A two-level trigger system selects the most interesting events
for use in the offline physics analyses. A more detailed description of the CMS detector can be found in Ref.~\cite{Chatrchyan:2008zzk}.

\section{Signal definition}
\label{sec:Signal_extraction}

\begin{figure}[b]
\begin{center}
\includegraphics[width=0.33\textwidth,angle=0]{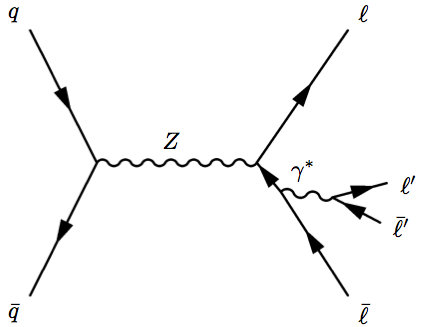}
\includegraphics[width=0.30\textwidth,angle=0]{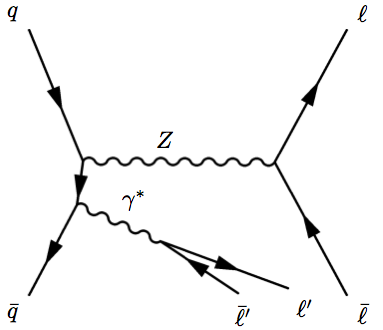}
\\ \caption{
  (Left) Diagram of the $\mathrm{Z}\rightarrow 4\ell$ process.
  (Right) Diagram of the $\mathrm{Z}\gamma^* \rightarrow 4\ell$ process for the irreducible background of
  $Z \to 2\ell$ production with the initial-state radiation undergoing an internal conversion $\gamma^* \to 2\ell$.
  Both Z and $\gamma^*$ are present in all propagators.
  The choice of propagators shown in the figures corresponds to the dominant contributions in the phase space $80 < m_{4\ell} < 100$\GeV.
}  \label{fig:qqZ4l_and_qqZZ4l}
\end{center}
\end{figure}

The leading-order (LO) Feynman diagram for the Z production and decay into four leptons, $\mathrm{q \bar{q}} \to \mathrm{Z} \to 4\ell$,
is presented in Fig.~\ref{fig:qqZ4l_and_qqZZ4l} (left).
This process is sometimes referred to as single-resonant four-lepton production.

We define signal events as those with four leptons,
$\mathrm{e^+e^-e^+e^-}$ ($4\mathrm{e}$),  $\mu^+\mu^-\mu^+\mu^-$ ($4\mu$), or $\mathrm{e^+e^-}\mu^+\mu^-$ ($\mathrm{2e}2\mu$),
with four-lepton invariant mass satisfying $80 < m_{4\ell} < 100$\GeV and dilepton masses $m_{\ell\ell}>4$\GeV for all six possible pairings of leptons.
The lower limit on $m_{\ell\ell}$ reduces background
that rises rapidly as $m_{\ell\ell}$ decreases.  The
branching fraction and cross section reported here are
defined by these phase space restrictions.

Predicted partial widths and branching fractions for Z boson decays
to $4\mathrm{e}$, $4\mu$, and $2\mathrm{e}2\mu$ final states
are summarized in Table~\ref{tab:BR_CalcHEP}.  The results are
obtained at LO using
CalcHEP 3.2~\cite{CalcHEP}, which takes quantum mechanical interferences into account.
The partial width for the $2\mathrm{e}2\mu$ channel is different from twice the width in either the $4\mathrm{e}$ or $4\mu$
channel because decays to four leptons
of the same flavour involve additional Feynman diagrams with permutations of same-sign leptons.
We do not have a
sufficient number of events to measure the differences
between decay rates for the three four-lepton final states.
We therefore measure the overall branching fraction $\mathcal{B}(\mathrm{Z} \to 4\ell)$
and assume the  $4\mathrm{e}$, $4\mu$, and $2\mathrm{e}2\mu$ relative branching fractions predicted by theory (Table~1).
The main irreducible background is the process
$\mathrm{q \bar{q}} \to \mathrm{Z} \gamma^* \rightarrow 4\ell$, for which the LO Feynman diagram is shown in Fig.~\ref{fig:qqZ4l_and_qqZZ4l} (right).
Events are simulated with the next-to-leading-order generator \POWHEG~\cite{powheg}. The effects of multiple pp collisions within each bunch crossing are taken into account in all simulated samples.

\begin{table}[h]
\centering
\begin{small}
    \topcaption{Partial widths and branching fractions for Z boson decays
    to $4\mathrm{e}$, $4\mu$, $2\mathrm{e}2\mu$ final states with  $m_{\ell\ell}>4$~GeV for all lepton pairs.
    The branching fractions are calculated with CalcHEP 3.2~\cite{CalcHEP} at LO using the total Z boson width $\Gamma_{\mathrm{tot}}=2.4952$~GeV~\cite{PDG}.
    Theoretical uncertainties are smaller than experimental uncertainties and are not shown in the table.
    }
    \begin{tabular}{lcccc}
\hline
Quantity of interest                                     &  $4\mathrm{e}$                  &  $4\mu$             &  $2\mathrm{e}2\mu$                &  $4\ell$   \\
\hline
\hline
                                             Partial width, $\Gamma_i$ (keV)                          & 2.95                   & 2.95      & 5.21       & 11.12     \\
Branching fractions, $\Gamma_i / \Gamma_{\mathrm{tot}}$              & $1.18 \times 10^{-6}$   & $1.18 \times 10^{-6}$   & $2.09 \times 10^{-6}$   & $4.45 \times 10^{-6}$ \\
                                             Relative fractions, $f_i = \Gamma_i / \Gamma_{4\ell}$    & 0.2655   & 0.2655   & 0.4690   &    -       \\
\hline
    \end{tabular}
    \label{tab:BR_CalcHEP}
\end{small}
\end{table}

\section{Signal extraction}

\label{sec:SigExt}

The trigger and selection criteria used in this analysis closely follow the $\mathrm{H} \to \mathrm{ZZ} \to 4\ell$ search by CMS~\cite{HIG-11-025}.
We use data collected with dielectron and dimuon triggers
selecting events with at least two electrons or two muons with transverse momentum $p_{T}>17$ and $8$~\GeV.
To match the trigger selection,
we require that at least two leptons reconstructed offline have
$p_{T}>20$ and $10$~\GeV. In this phase-space region, we expect and observe a high trigger efficiency
of 96--99\%, depending on the final state.

Muon candidates are reconstructed using two algorithms, one in
which tracks in the silicon strip tracker are matched to 
hits in
the muon detectors, and another in which a combined fit is performed to
signals in both the silicon strip tracker and the muon
system~\cite{CMS-PAS-MUO-10-002}.  The muon candidates are required to be successfully reconstructed by both
algorithms.
Other identification criteria (the number of measurements in the tracker and muon systems, the fit quality of the muon track, and the consistency of the track with the primary vertex) are also imposed
on the muon candidates to reduce the misidentification rate.
The vertex with the largest value of $\sum p_{T}^2$ for the associated
tracks is chosen to be the primary vertex.
According to simulation, this requirement provides the correct assignment for the primary
vertex in more than 99\% of both signal and background events. The average number of reconstructed vertices per event in 2011 data is approximately 7.

Electron reconstruction also involves two
algorithms, one in which energy clusters
in the electromagnetic calorimeter are matched to signals in the silicon pixel detector and another
in which tracks in the silicon strip tracker are matched to ECAL
clusters~\cite{CMS-PAS-EGM-10-004}. The electron candidates used in the analysis are
reconstructed by either algorithm.
Additional identification criteria (the spatial distribution of the shower in the ECAL, matching of the electron track with a cluster in the ECAL, and consistency of the track with the primary vertex) are imposed
on the electron candidates to reduce the misidentification rate.
Requirements are imposed on the geometrical properties of the electron track with respect to the primary vertex and neighbouring tracks, as well as requirements on the number of missing hits in the innermost tracker layers in order to remove electrons produced by photon conversions in the detector material.

Electrons are required to satisfy $p_{T} > 7$ GeV and $|\eta| < 2.5$, while muons are required to satisfy  $p_{T} > 5$ GeV and $|\eta| < 2.4$.

Leptons produced in the decay of Z  (or $\gamma^{*}$) bosons are
typically isolated from hadronic activity in the event. The scalar sum
of transverse energy deposits in the calorimeters and the
transverse momenta of tracks in a cone
of radius
$\Delta R = \sqrt{(\Delta \phi)^2 + (\Delta \eta)^2} = 0.3$
in $\eta$-$\phi$ space around each lepton is computed. This sum is  corrected for the contribution from the other lepton candidates as well as for the average hadronic activity in an event.
The ratio of this corrected sum to the lepton $p_{T}$ is required to be
smaller than 0.275.

To suppress background events with leptons originating from secondary
vertices (such as heavy quark decays or misidentified pions),  the
significance of the impact parameter of each lepton relative to
the primary vertex, $s_\mathrm{3D} = d_\mathrm{3D} / \sigma_\mathrm{3D}$, is required to satisfy $|s_\mathrm{3D}|<4$, where $d_\mathrm{3D} $ is the impact
parameter in three dimensions and $\sigma_\mathrm{3D}$ is its associated
uncertainty.

Events are selected with at least four leptons
satisfying the above criteria.
The same-flavour, opposite-charge lepton pair with mass
closest to the nominal \cPZ\ boson mass is  identified.
A second pair is constructed from the remaining leptons
with the highest $p_{T}$ satisfying the requirement
of the same flavour and opposite charge.
We require all six dilepton combinations
constructed from the four selected leptons
to satisfy $m_{\ell\ell} > 4\GeV$ and the four-lepton invariant mass must be in the range $80 < m_{4\ell} < 100$~\GeV. The overall reconstruction and selection efficiencies for simulated signal events are found to be 25\%, 59\%, and 33\% for the $4\mathrm{e}$, $4\mu$ and $2\mathrm{e}2\mu$ channels, respectively. The overall theoretical acceptance efficiencies (per channel, in the same order) are 8.9\%, 14\%, and 11.1\%.

We estimate the irreducible four-lepton background from simulation using selection efficiencies
corrected for differences between the simulated efficiencies and those measured in data.

Processes with one or more misidentified or nonprompt leptons contributing to
the signal region $80 < m_{4\ell} < 100$~\GeV are dominated by $\mathrm{Z}+\mathrm{X}$ production.
Their rate is measured in data using control regions and is found to be negligible.
The methodology used is identical to that used in the $\mathrm{H} \to \mathrm{ZZ} \to 4\ell$ search~\cite{CMSobsJul2012}.

We use $\mathrm{pp} \to \mathrm{Z} \to 2\mu$ events as a control sample in order to
cancel systematic uncertainties in the measurement of the branching fraction
$\mathcal{B}(\mathrm{Z} \to 4\ell)$.
The muon selection criteria are identical to those in the $\mathrm{Z} \to 4\ell$ event selection
with $p_{T}$ thresholds $p_{T} > 20$ and 10\GeV applied to the higher and lower $p_T$ muons, respectively.

\section{Results}
\label{sec:results}

\subsection{Observation of \texorpdfstring{$\mathrm{Z} \to 4\ell$}{Z to 4 l} decays}

\begin{figure}[b]
\begin{center}
   \includegraphics[width=0.55\textwidth,angle=0]{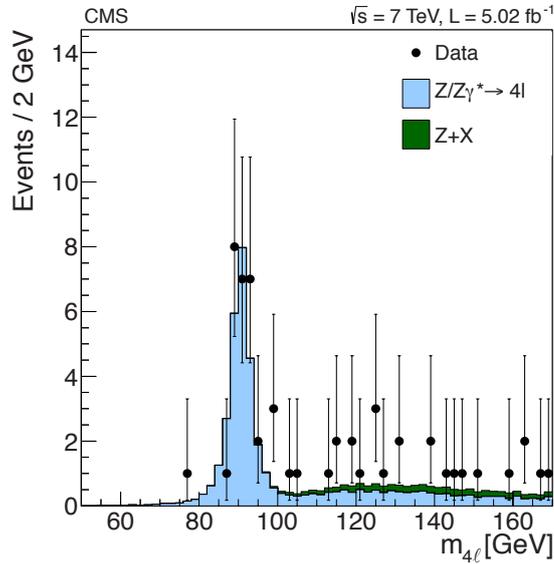}
   \\ \caption{Four-lepton invariant mass distribution for events passing all selection requirements except that on $m_{4\ell}$.
   The data are shown by points. The
   filled histograms represent standard model expectations for $\mathrm{pp} \to \mathrm{Z}/\mathrm{Z}\gamma^{*} \to 4\ell$ 
   and for reducible backgrounds.    
   The three final states, $4\mathrm{e}$, $4\mu$, and $2\mathrm{e}2\mu$, are combined.
   }
   \label{fig:m4l_data_MC}
\end{center}
\end{figure}

Figure~\ref{fig:m4l_data_MC} shows the four-lepton mass distribution for events
that pass all the selection criteria except the invariant
mass requirement.  Events with all three final states,
$4\mathrm{e}$, $4\mu$, and $2\mathrm{e}2\mu$, are included.   The prominent peak at
$m_{4\ell} = m_Z$ constitutes the first observation of
$\mathrm{pp} \to \mathrm{Z} \to 4\ell$.  The number of events with $80 < m_{4\ell} < 100$\GeV is \Nobs, in good agreement with the standard model expectation
of \nTOTexp\ events.  Table~\ref{tab:events} shows good agreement between the
expected and observed events in each of the $4\mathrm{e}$, $4\mu$, and $2\mathrm{e}2\mu$ channels.  The probability for a background fluctuation
to be at least as large as
the observed maximum excess can be evaluated by generating sets of
simulated data incorporating all statistical and systematic
uncertainties. This probability can be expressed as a corresponding
number of standard deviations using the one-sided Gaussian tail
convention.  The results presented
in this paper are obtained using asymptotic formulae~\cite{LHC-HCG-report} and include the systematic uncertainties
described in the next subsection.  The statistical
significance of the peak in Fig.~\ref{fig:m4l_data_MC} is \Significance .

\begin{table}[h]
\centering
\begin{small}
    \topcaption{Numbers of expected and observed events with $80 < m_{4\ell} < 100$\GeV.  The yields are determined by means of fits described in the text; the fit results in the $4\mathrm{e}$ channel are not meaningful because only two events are selected.
    }
    \begin{tabular}{lcccc}
\hline
Final state channels                           &  $4\mathrm{e}$     &  $4\mu$    & $2\mathrm{e}2\mu$  &  $4\ell$      \\
\hline
\hline
Irreducible background ($\mathrm{pp} \to \mathrm{Z} \gamma^* \to 4\ell$) &    0.07   &  0.25      &  0.14    &    $0.46\pm 0.05$     \\
Other (reducible) backgrounds                    &    0.01   &  0.01      &  0.05    &    $0.07\pm0.1$   \\
Expected signal ($\mathrm{pp} \to \mathrm{Z} \to 4\ell$)         &    3.8    &  13.6      &  12.0     &   $29.4\pm 2.6$    \\
\hline
Total expected (simulation)                                &    3.9    & 13.9       &  12.2     &   $30.0\pm 2.6$      \\
\hline
Observed events                                     &     2     &  14        &   12     &    28    \\
\hline
\hline
Yield from  fit to the observed mass distribution                   &      -      &  $13.6\pm3.8$      &  $11.5\pm3.1$     &   $27.3\pm5.4$     \\
\hline
    \end{tabular}
    \label{tab:events}
\end{small}
\end{table}

\subsection{Measurement of \texorpdfstring{$\sigma(\mathrm{pp} \to \mathrm{Z}) \,  \mathcal{B}(\mathrm{Z} \to 4\ell)$ and $\mathcal{B}(\mathrm{Z} \to 4\ell)$}{sigma(pp to ZZ) B(Z to 4 l) and B(Z to 4 l)}}

The numbers of $\cPZ \to 4\ell$ (signal) and $\cPZ \to 2\mu$ (control) events
passing the selection may be expressed as follows:

\begin{eqnarray}
N^{4\ell} - N^{4\ell}_\mathrm{bkg} &=& \sum_{i} L\, \sigma(\Pp\Pp \to \cPZ)\, \mathcal{B}(\cPZ \to 4\ell)\, f_i\, \epsilon^\text{acc}_i\, \epsilon^\text{exp}_i\, c_i \ \ \ (i=4\Pe,4\mu,2\Pe 2\mu),
\label{eq:N4l}\\
N^{2\mu} - N^{2\mu}_\mathrm{bkg} &=& L\, \sigma(\Pp\Pp \to \cPZ)\, \mathcal{B}(\cPZ \to 2\mu)\, \epsilon^\text{acc}_{\cPZ \to 2\mu}\, \epsilon^\text{exp}_{\cPZ \to 2\mu}\, c_{\cPZ \to 2\mu},
\label{eq:N2mu}
\end{eqnarray}

where

\begin{itemize}
\item $L$ denotes the integrated luminosity;
\item $\sigma(\Pp\Pp \to \cPZ) = 26.9\unit{nb}$ is the theoretical \cPZ\ boson production cross section
($80 < m_{4\ell}  < 100\GeV$), calculated with \textsc{fewz}~\cite{Gavin:2010az} at NNLO level;
\item $\mathcal{B}(\cPZ \to 4\ell)$ is the signal decay branching fraction (for $m_{\ell\ell}>4\GeV$);
\item $\mathcal{B}(\cPZ \to 2\mu) = 0.03366\pm0.00002$ is the $\mathrm{Z} \to 2\mu$ branching fraction~\cite{PDG};
\item $f_i$ is the relative fraction of all $4\ell$ events in the $i$th channel ($i = 4\Pe, 4\mu, 2\Pe 2\mu$) from Table~\ref{tab:BR_CalcHEP};
\item $\epsilon^\mathrm{acc}_i$ is the theoretical acceptance of the lepton $\pt$ and $\eta$ requirements used in the analysis;
\item $\epsilon^\mathrm{exp}_i$ is the experimental efficiency, as obtained from the simulation, to reconstruct events within the acceptance;
\item $c_i$ is the data-to-simulation correction factor for the experimental efficiency derived from Monte Carlo events.
The $c_i$ are within one percent of unity.
\end{itemize}

Equation~(\ref{eq:N4l}) allows the extraction of the production cross section times branching fraction $\sigma(\mathrm{pp} \to \mathrm{Z})\,\mathcal{B}(\mathrm{Z} \to 4\ell) = 112^{+23}_{-20} \stat\unit{fb}$,
while the ratio of Eqns.~(\ref{eq:N4l}) and (\ref{eq:N2mu}) allows extraction of the branching fraction $\mathcal{B}(\mathrm{Z} \to 4\ell) = \left( 4.2^{+0.9}_{-0.8}\,(\text{stat.}) \right) \times 10^{-6}$ with cancellation of several systematic uncertainties.

The signal and the irreducible background have the following {\color{black}systematic} uncertainties:

\begin{itemize}

\item Interference between the signal and background production is 
less than 0.2\% in the range $80 < m_{4\ell} < 100$\GeV, and is accounted for as a source of systematic uncertainty.

\item Uncertainties in parton distribution functions lead to uncertainties
in the total cross section and acceptance. For $\mathrm{pp} \to \mathrm{Z} \gamma^* \to 4\ell$
and $\mathrm{pp} \to \mathrm{Z} \to 2\mu$, these uncertainties are 4\%~\cite{HIG-11-025}
and 1\%~\cite{EWK-10-005}.
For $\mathrm{pp} \to \mathrm{Z} \to 4\ell$ we obtain an uncertainty of 4\% by following the same procedure as in Ref.~\cite{HIG-11-025}.

\item
The theoretical uncertainties related to the choice of QCD renormalization/factorization scales are evaluated by varying these scales up and
down by a factor of
two. For $\mathrm{pp} \to \mathrm{Z} \gamma^* \to 4\ell$ and $\mathrm{pp} \to \mathrm{Z} \to 2\mu$, these
uncertainties are both 2\%~\cite{HIG-11-025,EWK-10-005}.
For $\mathrm{pp} \to \mathrm{Z} \to 4\ell$ we obtain an uncertainty of 1\% by following the same procedure as in Ref.~\cite{HIG-11-025}.

\item The effect of  bremsstrahlung radiation on event acceptance is estimated to be negligible.

\item Lepton reconstruction, isolation, and impact parameter uncertainties as
well as uncertainties associated with the high-level trigger efficiency
are determined using a ``tag-and-probe" method~\cite{EWK-10-005} where efficiencies
and uncertainties are measured in samples of  $\mathrm{Z} \to 2\ell$ events.
The statistical uncertainties
on the lepton efficiencies obtained in different ($p_T$, $\eta$) bins
are propagated to four-lepton reconstruction uncertainties for
$\mathrm{Z} \to 4\ell$ and $\mathrm{Z} \gamma^* \to 4\ell$ events.
These uncertainties range from 1\% to 6\%. Correlation between the ``tag-and-probe" efficiency measurement at the $\mathrm{Z} \to 2\ell$ calibration signal are taken into account.

\item The uncertainty on the integrated luminosity is
2.2\%~\cite{lumiPAS}.

\end{itemize}

The reducible background is estimated directly from data.
The predicted yield of the reducible backgrounds is approximately 0.1 event (less than one percent of the signal yield,
as shown in Table~\ref{tab:events}) and has a total uncertainty, also evaluated from
data, of 130\%.

To  account for systematic uncertainties,
including  correlations among different channels, as well as between signal and background,
we construct the  likelihood for the four observed final states ($4\mathrm{e}$, $4\mu$, $2\mathrm{e}2\mu$, and $2\mu$).
Using the full likelihood in conjunction with the profile
likelihood method~\cite{LHC-HCG-report},
we measure the production cross section times branching fraction and the branching fraction for $\mathrm{Z} \to 4\ell$ to be

\begin{eqnarray*}
\sigma(\pp\to\mathrm{Z}) \, \mathcal{B}(\mathrm{Z} \to 4\ell) &=& 112^{+23}_{-20}\,\text{(stat.)} ^{+7}_{-5}\text{(syst.)} ^{+3}_{-2}\,\text{(lum.)}\unit{fb},\\
\mathcal{B}(\mathrm{Z} \to 4\ell) &=& \left( 4.2^{+0.9}_{-0.8}\,\text{(stat.)} \pm 0.2\,\text{(syst.)} \right) \times  10^{-6}.
\end{eqnarray*}

These measurements agree with the standard model predictions, $120\pm 5$\unit{fb} and $4.45 \times 10^{-6}$, respectively.

\subsection{\texorpdfstring{$\mathrm{Z} \to 4\ell$ decays as a calibration channel in the $\mathrm{H} \to \mathrm{ZZ} \to 4\ell$ search}{Z to 4 l decays as a calibration channel in the H to ZZ to 4 l search}}

\begin{figure}[h]
\begin{center}
   \includegraphics[width=0.49\textwidth,angle=0]{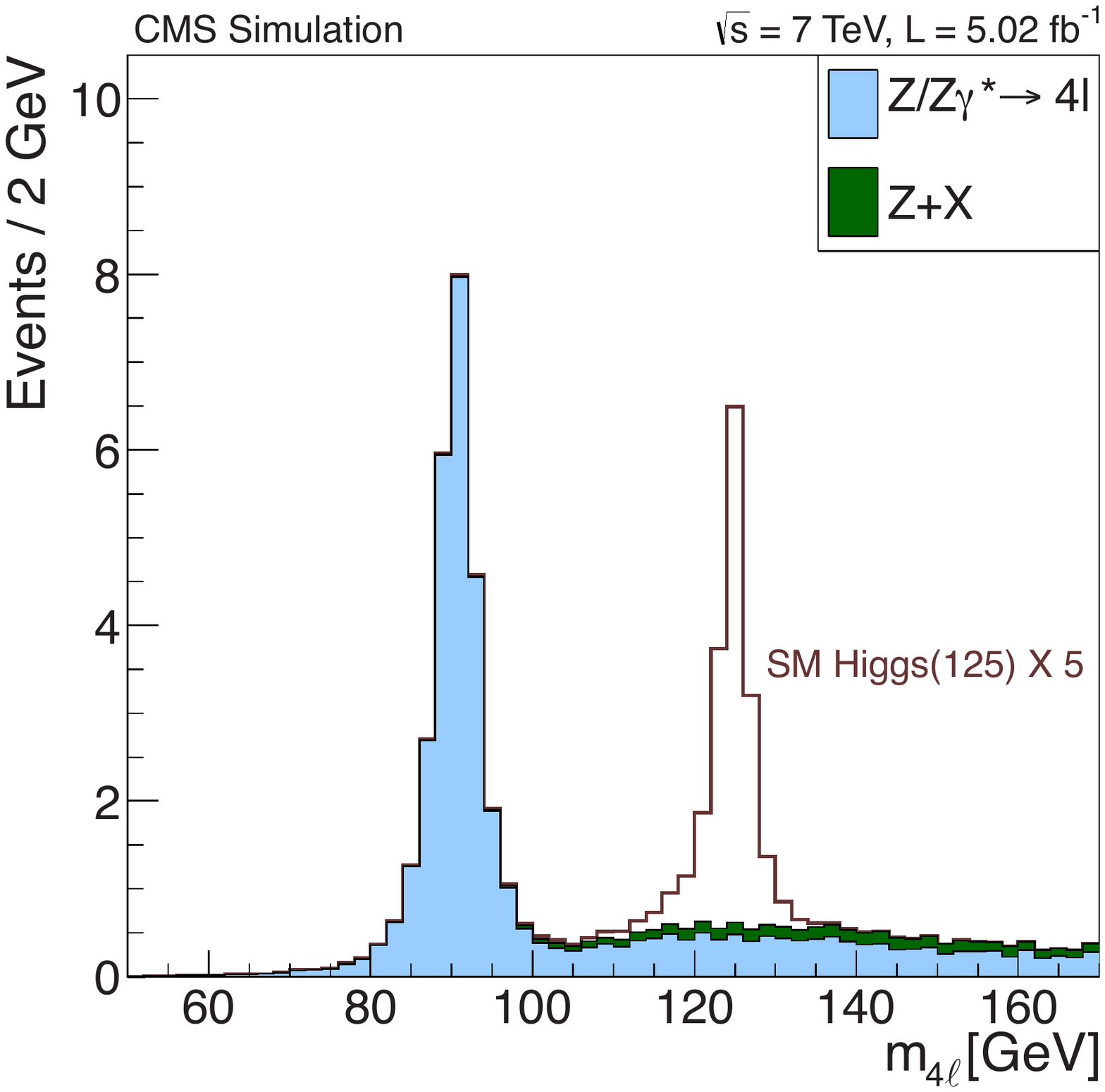}
   \includegraphics[width=0.49\textwidth,angle=0]{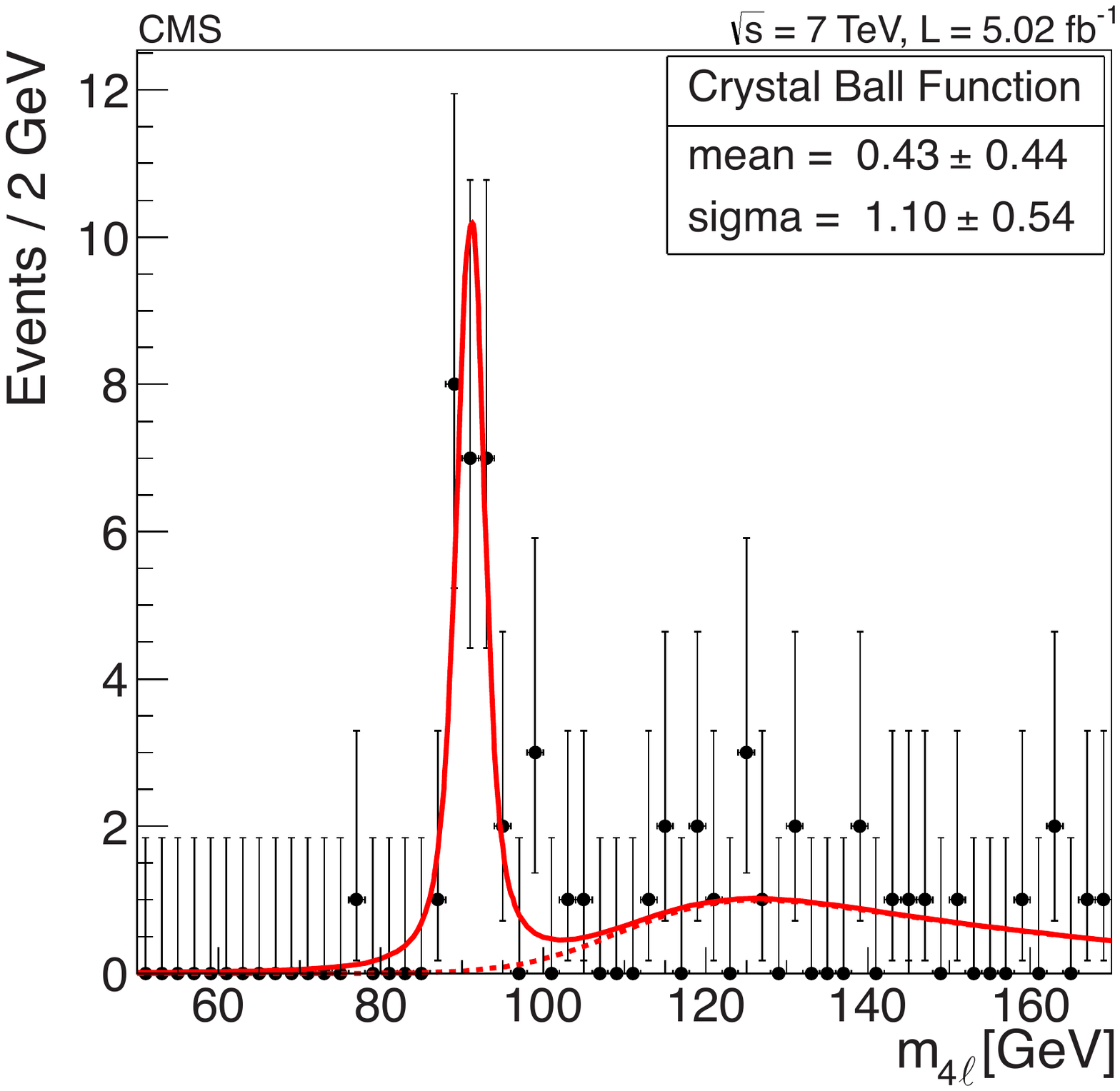}
   \\ \caption{
   (Left) Four-lepton mass distribution in simulation for $\mathrm{pp} \to 4\ell$, without the Higgs boson (light shaded histogram), $\mathrm{Z}+\mathrm{X}$ background (dark
 shaded histogram),  and
   $\mathrm{pp} \to \mathrm{H} \to \mathrm{ZZ} \to 4\ell$ for a Higgs boson mass $m_{\mathrm{H}}=125$\GeV. The three contributions are stacked.
  The standard model cross section for the Higgs boson is scaled up by a factor of 5.
   (Right) Four-lepton mass distribution with data represented by the points with error bars.
   The three final states, $4\mathrm{e}$, $4\mu$, and $2\mathrm{e}2\mu$, are combined.
   The solid line represents a simultaneous fit to the background
   and Z boson peak.
   }
   \label{fig:m4l_data_noMC}
\end{center}
\end{figure}

The $\mathrm{Z} \to 4\ell$ decays give a narrow resonant peak in the
four-lepton invariant mass distribution, which can be used as a
calibration channel in the context of the Higgs boson search in the $\mathrm{H}
\to \mathrm{ZZ} \to 4\ell$ decay mode.
Figure~\ref{fig:m4l_data_noMC} (left) shows that the number of events in the $\mathrm{Z} \to
4\ell$ peak at $m_{4\ell} = m_Z$ is approximately an order of magnitude larger than the
expected number of events for the SM Higgs boson with a mass $m_{\mathrm{H}} = 125$~\GeV~\cite{CMSobsJul2012,ATLAS:2012gk}.
The $\mathrm{Z}
\to 4\ell$ peak can be used for a direct calibration of the four-lepton
mass scale, the four-lepton mass resolution, and the overall
four-lepton reconstruction efficiency in a  phase space similar to
the Higgs boson four-lepton decays. Such a direct calibration using
the $\mathrm{Z} \to 4\ell$ peak is complementary to the current
indirect tag-and-probe method making use of $\mathrm{Z} \to 2\ell$ events,
with the benefit that $\mathrm{Z} \to 4\ell$ has similar kinematics to $\mathrm{H} \to \mathrm{ZZ} \to 4\ell$.

Figure~\ref{fig:m4l_data_noMC} (right) shows the results of an unbinned maximum-likelihood
fit to the four-lepton mass distribution for the observed events.
The fitted signal yield is shown in the last line of Table~\ref{tab:events},
along with the results of separate fits to the mass distributions
in the three channels.
The background shape comes from a simulation of $\mathrm{pp} \to \mathrm{ZZ} \to 4\ell$ events
with the overall normalization floating in the fit.
Differences in shape between the reducible and irreducible backgrounds are ignored because of the small contribution of the former.
The signal shape is a convolution of the Breit--Wigner and Crystal Ball~\cite{CBfunction}
functions. The central value and width of the Breit--Wigner function are fixed at
the Z boson mass and width~\cite{PDG}. The Crystal Ball parameters
are free to vary in the fit. Data used for this analysis is only a (lower energy) subset of that used in the recent discovery paper~\cite{CMSobsJul2012}.

From the fit results, as shown in Fig.~\ref{fig:m4l_data_noMC} (right), one can see that the offset of the peak is $0.4 \pm 0.4\GeV$ (to the right)
(the ``mean" value of the Crystal Ball function represents an offset with respect to the fixed Breit--Wigner function peak position),
or, in relative
units, $0.4 \pm 0.4\%$. These numbers can be used to constrain the possible systematic uncertainty
of the four-lepton mass scale. With the current data there is no evidence for a statistically significant bias.

The Crystal Ball width (sigma of the Gaussian core) returned by the fit is $1.1 \pm 0.5\GeV$, consistent with the simulation-based expectations of
1.0\% for $4\mu$ and 1.4\% for $2\mathrm{e}2\mu$ (which for a Higgs boson with mass $125\GeV$ corresponds to $1.3\GeV$ and $1.8\GeV$) for a Higgs signal in this  mass range.
There are only two $4\mathrm{e}$ events, so the mass resolution of $4\mathrm{e}$ events cannot be constrained.
With the current data, we can measure the average four-lepton mass resolution
with about 45\% statistical uncertainty. (For completeness, the other two parameters of the Crystal Ball function from the fit are $N = 0.4 \pm 1.7$ and $\alpha = 12 \pm 10$.)

\section{Summary}
The first observation of the Z boson decaying to four leptons (electrons and/or muons) in proton-proton collisions has been presented.
The data set analyzed corresponds to an integrated luminosity of 5.02\fbinv at $\sqrt{s}=7$\TeV.
A pronounced resonance peak, with a statistical significance of 9.7$\,\sigma$,
is observed in the distribution
of the invariant mass of four leptons with  mass and
width consistent with expectations for Z boson decays.
The event yields,
branching fraction, and cross section reported here are defined by
phase space restrictions on the leptons, namely, $80 < m_{4\ell} < 100$\GeV, where $m_{4\ell}$ is the invariant mass of the
four leptons, and $m_{\ell\ell} > 4$\GeV for all pairs of leptons, where $m_{\ell\ell}$ is the two-lepton invariant mass.
We observe \Nobs ~events, in agreement
with the expectation of \nTOTexpErr ~events,
comprised of \nZexpErr ~ $\mathrm{Z} \to 4\ell$ events and
\nBKGDexpErr ~events from backgrounds.
The measured branching fraction
is \BR , in agreement with the standard model prediction of $4.45 \times 10^{-6}$.
The measured cross section times branching fraction is \XSxBR ,
also consistent with the standard model prediction of 120$\,$fb.
The four-lepton mass peak arising from $\mathrm{Z} \to 4\ell$ decays
provides a calibration channel for the Higgs boson search
in the $\mathrm{H} \to \mathrm{ZZ} \to 4\ell$ decay mode.

\section*{Acknowledgements}

We congratulate our colleagues in the CERN accelerator departments for the excellent performance of the LHC machine. We thank the technical and administrative staff at CERN and other CMS institutes, and acknowledge support from BMWF and FWF (Austria); FNRS and FWO (Belgium); CNPq, CAPES, FAPERJ, and FAPESP (Brazil); MES (Bulgaria); CERN; CAS, MoST, and NSFC (China); COLCIENCIAS (Colombia); MSES (Croatia); RPF (Cyprus); MoER, SF0690030s09 and ERDF (Estonia); Academy of Finland, MEC, and HIP (Finland); CEA and CNRS/IN2P3 (France); BMBF, DFG, and HGF (Germany); GSRT (Greece); OTKA and NKTH (Hungary); DAE and DST (India); IPM (Iran); SFI (Ireland); INFN (Italy); NRF and WCU (Korea); LAS (Lithuania); CINVESTAV, CONACYT, SEP, and UASLP-FAI (Mexico); MSI (New Zealand); PAEC (Pakistan); MSHE and NSC (Poland); FCT (Portugal); JINR (Armenia, Belarus, Georgia, Ukraine, Uzbekistan); MON, RosAtom, RAS and RFBR (Russia); MSTD (Serbia); SEIDI and CPAN (Spain); Swiss Funding Agencies (Switzerland); NSC (Taipei); ThEP, IPST and NECTEC (Thailand); TUBITAK and TAEK (Turkey); NASU (Ukraine); STFC (United Kingdom); DOE and NSF (USA).

\bibliography{auto_generated}   

\cleardoublepage \appendix\section{The CMS Collaboration \label{app:collab}}\begin{sloppypar}\hyphenpenalty=5000\widowpenalty=500\clubpenalty=5000\textbf{Yerevan Physics Institute,  Yerevan,  Armenia}\\*[0pt]
S.~Chatrchyan, V.~Khachatryan, A.M.~Sirunyan, A.~Tumasyan
\vskip\cmsinstskip
\textbf{Institut f\"{u}r Hochenergiephysik der OeAW,  Wien,  Austria}\\*[0pt]
W.~Adam, E.~Aguilo, T.~Bergauer, M.~Dragicevic, J.~Er\"{o}, C.~Fabjan\cmsAuthorMark{1}, M.~Friedl, R.~Fr\"{u}hwirth\cmsAuthorMark{1}, V.M.~Ghete, J.~Hammer, N.~H\"{o}rmann, J.~Hrubec, M.~Jeitler\cmsAuthorMark{1}, W.~Kiesenhofer, V.~Kn\"{u}nz, M.~Krammer\cmsAuthorMark{1}, D.~Liko, I.~Mikulec, M.~Pernicka$^{\textrm{\dag}}$, B.~Rahbaran, C.~Rohringer, H.~Rohringer, R.~Sch\"{o}fbeck, J.~Strauss, A.~Taurok, P.~Wagner, W.~Waltenberger, G.~Walzel, E.~Widl, C.-E.~Wulz\cmsAuthorMark{1}
% ...
\vskip\cmsinstskip
\dag:~Deceased\\
1:~~Also at Vienna University of Technology, Vienna, Austria\\
% ...

\end{sloppypar}
\end{document}